\documentclass[twocolumn,prl]{revtex4}
\usepackage{graphics,graphicx}
\begin{document}
\title{Anomalous magnetoresistance effect in nanoengineered material}
\author{S.~Dickert$^{1}$}
\author{D.K.~Singh$^{2,3,*}$}
\author{R.~Thantirige$^{1}$}
\author{M. T.~Tuominen$^{1,*}$}
\affiliation{$^{1}$University of Massachusetts, Amherst, MA 01002, USA}
\affiliation{$^{2}$National Institute of Standard and Technology, MD 20899, USA}
\affiliation{$^{3}$Department of Materials Science and Engineering, University of Maryland, College Park, MD 20742,USA}
\affiliation{$^{*}$email: dsingh@nist.gov; tuominen@physics.umass.edu}

\maketitle

\textbf{The periodic response of magnetoresistance to an externally tunable parameter, such as magnetic field or chemical composition, in the bulk or an artificially designed material has been exploited for technological applications\cite{Chappert} as well as to advance our understanding of many novel effects of solid state physics.\cite{Rosenbaum,Chopra,Siegert} Some notable examples are the giant magnetoresistance effect in layered materials,\cite{Parkin} the quantum hall effect in semiconductor heterostructure\cite{Zhang} and the phase coherence of electronic wave function in disordered metals\cite{Webb}. In recent years, the ability to engineer materials at the nanoscale has played a key role in exploring new phenomenon. Using a system involving periodic Co dots array in direct contact with a surrounding polycrystalline Cu film, we report the observation of giant thermal hysteresis and an anomalous oscillatory magnetoresistance behavior. The unusual aspects of oscillatory magnetoresistance include its observation along only one field scan direction in an intermediate temperature range of 100 K $\leq$ $T$ $\leq$ 200 K. Reducing the thickness of the Cu film weakens the magnetoresistance oscillation. These properties suggest a new phenomenon, which could be harnessed for future technological applications.}

\begin{figure}
\centering
\includegraphics[width=9cm]{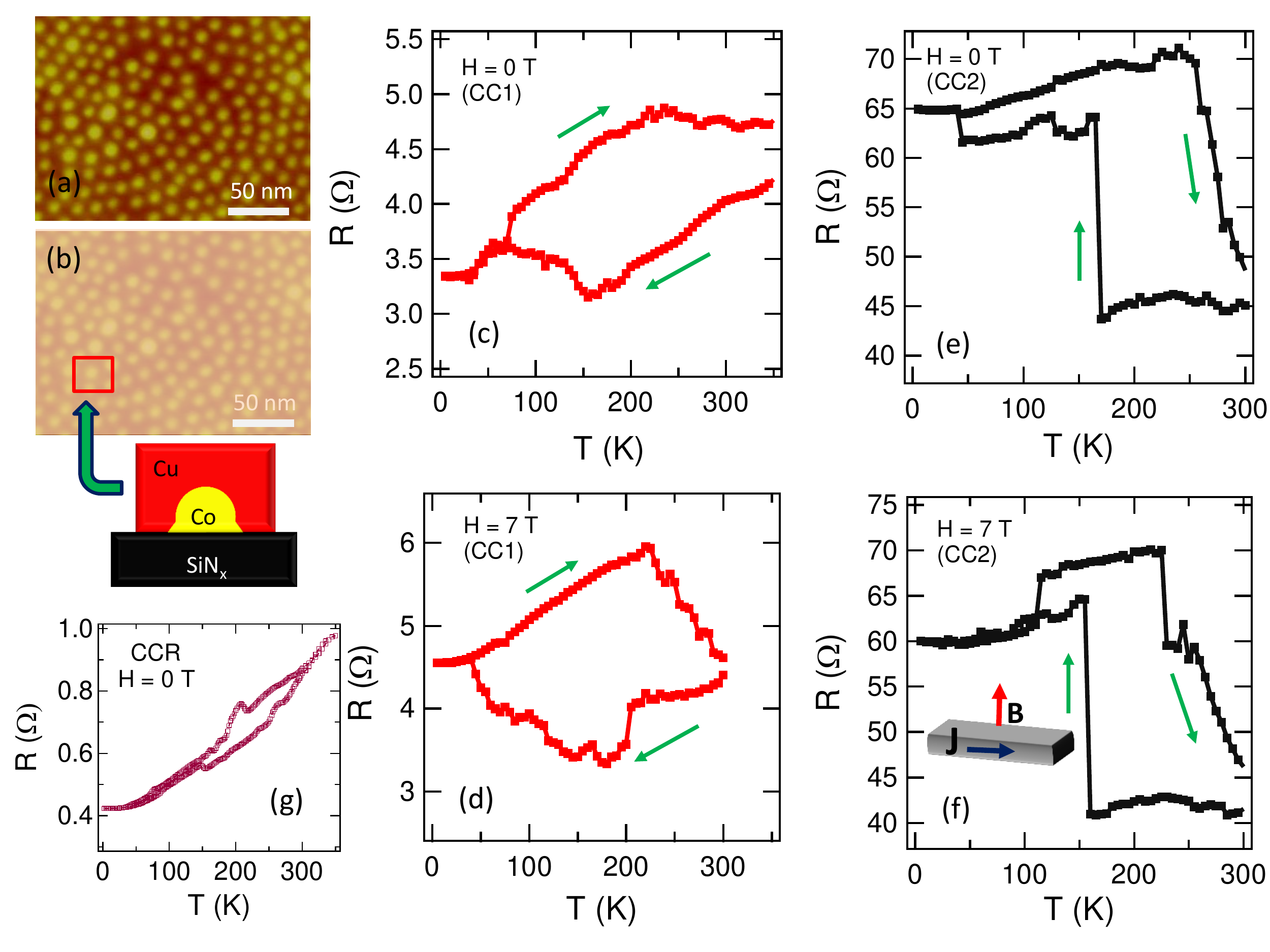} \vspace{-2mm}
\caption{\textbf{Schematic description of sample configuration and thermal characterization of nanoengineered materials. a,} Top view of cobalt dot array on a silicon nitride substrate. Atomic force micrograph shows that the magnetic dots form a locally hexagonal pattern. \textbf{b,} Schematic depiction of the cobalt dot array coated with a thin polycrystalline copper film (15-30 nm) after 15 minutes of argon plasma treatment. A three dimensional schematic of an individual cobalt dot surrounded by copper is highlighted by a small contour. \textbf{c-f,} Characteristic plots of electrical resistance measurements as a function of temperature at $H$ = 0 and 7 T of two nanoengineered systems (see text), CC1 and CC2 respectively. Magnetic field was applied perpendicular to the sample plane. \textbf{g,} Resistance versus temperature measurement of a control sample, CCR, where the cobalt particles are randomly distributed over the silicon nitride substrate and coated with 30 nm of copper film. Clearly, a much smaller thermal hysteresis in CCR signifies the importance of locally hexagonal topography of cobalt dots in nanoengineered materials.
} \vspace{-4mm}
\end{figure}

The multitude of physical effects reflected by the magnetoresistance (MR) oscillation as a function of an extrinsic parameter directly impacts the understanding of materials intrinsic properties. In many cases, this effect  is used to deduce the Fermi surface of a physical system,\cite{Sebastian} which is of fundamental importance in understanding the electronic properties of materials. In artificial materials, such as a micrometer size normal metal ring of silver or gold metal, the magnetoresistance oscillation was found to be directly related to the band structure properties of disordered materials.\cite{Cheung} About two decades ago, the observation of magnetoresistance oscillation in a multilayered Co films as a function of the thickness of the Cu layer spacer revolutionized the magnetic data storage technology.\cite{Parkin} In this letter, we report on using the combination of Cu and Co elements to nanoengineer a composite system via a straightforward fabrication scheme. The nanoengineered material consists of a periodic array of Co dots (12 nm in diameter and 3 nm in thickness, with a periodicity of 28 nm) in direct multidirectional  contact with encapsulating thin layer of polycrystalline Cu film (30 nm) (see Methods section). In rest of the manuscript, this sample configuration is called CC1. The topography of this nanostructured material is shown in Fig. 1a-b. As discussed below, in  electrical transport measurements on CC1, we have observed an anomalous quasi-periodic oscillation of magnetoresistance as a function of field. Under identical experimental conditions, we also fabricated two more samples: in one case, the cobalt dots of same thicknesses (3 nm) were randomly distributed across the substrate  (CCR) and encapsulated with 30 nm thick copper film, while the second sample (CC2) was identical to the sample CC1, except that the thickness of the copper film was reduced by half ($\simeq$ 15 nm).

The electrical measurements were performed on all three samples (CC1, CC2 and CCR) using four probe contact method, utilizing the silver paint to make equidistant point gold wire contacts along a line. In order to make reliable metallic contact between the sample and the measurement unit, the sample was cycled from 300 K to 350 K twice.  Electrical resistance measurements were performed between 350 K and 2 K. As shown in Fig. 1c-g, the resistance of samples with periodic (CC1 and CC2) and non-periodic (CCR) structure exhibit different thermal responses. A giant thermal hysteresis in the electrical measurements of samples CC1 and CC2 are observed. As the measurement temperature decreases, the electrical resistance of nanoengineered materials starts increasing around 150 K (albeit much more sharply in CC2), before decreasing again below 50 K (Fig. 1c). The cooling and warming resistivity curves separate above 50 K and do not retrace each other till the highest measurement temperature of 350 K (Fig. 1c). A thermal hysteresis in the electrical transport measurement can occur due to various reasons, as found in the colossal magnetoresistance device (CMR) or thin film of Vanadium oxide. The percolative transport through ferromagnetic domains in a CMR material\cite{Uehara} or, the development of biaxial strain as a result of lattice mismatch between the film and the substrate in thin film of vanadium oxide\cite{Natale}are often cited as the primary reason of thermal hysteresis in these materials. In polycrystalline thin films, the effect of strain at the interfaces of materials or thin film and substrate due to the differences in the crystal lattice parameters and thermal expansion coefficients is particularly significant. The resulting stress affects the phase transformation and the physical properties of thin film systems as a function of temperature, which ultimately leads to a thermal hysteresis.\cite{Podzorov} A similar mechanism leading to the giant thermal hysteresis in this case cannot be ruled out. We also performed the resistance versus temperature measurements in magnetic field, applied perpendicular to the sample plane  at $T$ = 300 K, at many field values between 0 and 7 T. In Fig.1d and 1f, we have plotted representative measurements at $H$ = 7 T for CC1 and CC2 samples, respectively. As we see in these figures, the application of magnetic field clearly affects the thermal hysteresis behavior. In fact, even a small field application (see supplementary materials) changes the thermal hysteresis properties. Therefore, the strain development, at the interface of Co and Cu polycrystals or between the substrate and the polycrystalline materials, alone may not be the likely cause of large thermal hysteresis in the nanoengineered materials. A very weak thermal hysteresis in the non-periodic sample, CCR, signifies the locally hexagonal topography of Co dots. The weak thermal hysteresis in CCR can possibly be attributed to the random distribution of large Co clusters, which minimizes or cancels the strain effect between Co and Cu materials. 

\begin{figure}
\centering
\includegraphics[width=8.5cm]{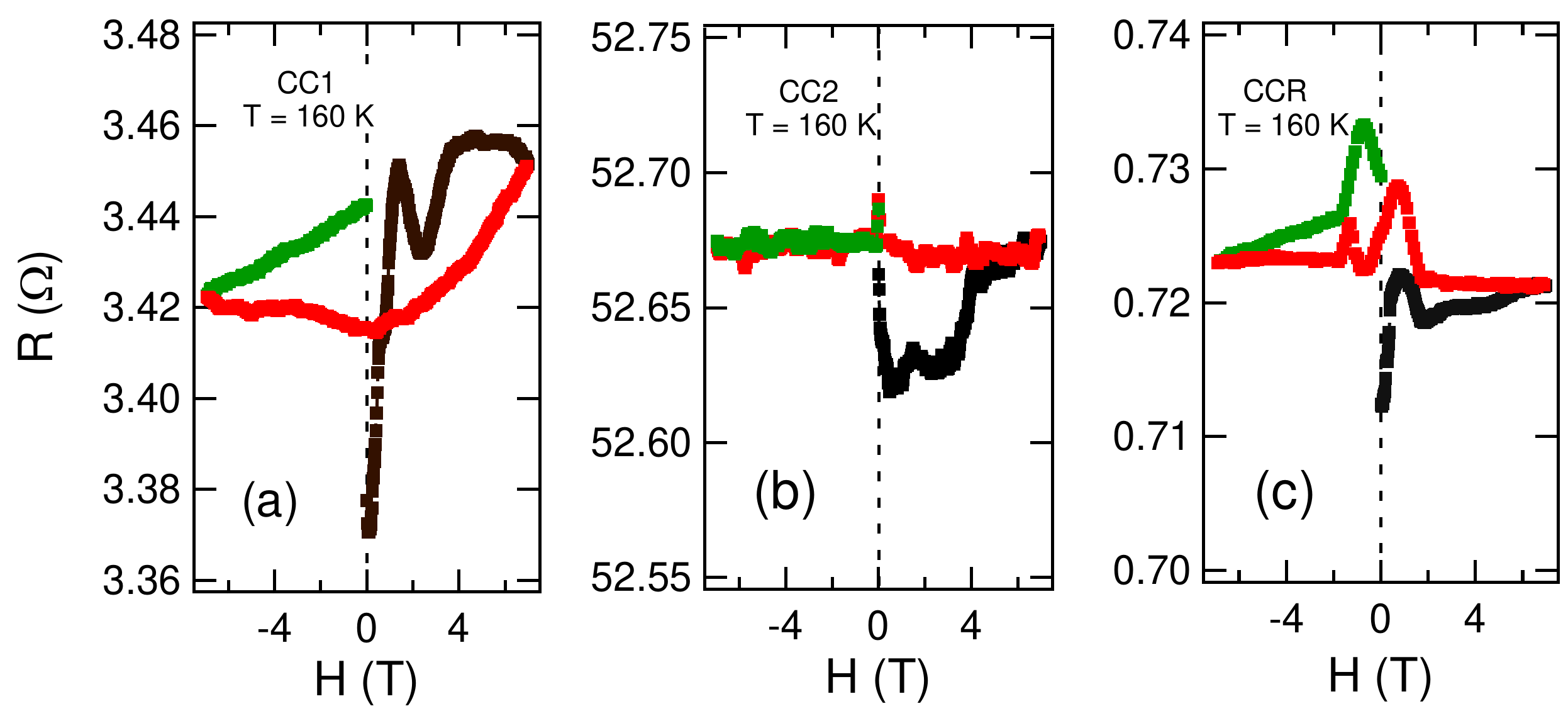} \vspace{-2mm}
\caption{\textbf{Resistance versus magnetic field measurements at T = 160 K after cooling in the zero field}. \textbf{a,} An oscillatory behavior in the resistance ($R$) versus field ($H$) measurement of CC1 is observed along the first field scan direction. Black markers represent the first scan from 0 to 7 T field. Red markers represent the field scan from 7 T to -7 T and green markers represent the field scan from -7 T to 0 T. A thermal procedure was followed to observe the oscillatory behavior (see text). \textbf{b,} Magnetoresistance measurement of CC2 also exhibits the oscillatory behavior, perhaps weaker, in one field scan direction. \textbf{c,} In CCR, a set of symmetric peaks in both field scan directions is observed.
} \vspace{-4mm}
\end{figure}

In an interesting observation in the electrical measurements on nanoengineered samples, particularly in CC1, quasi-oscillatory behavior in magnetoresistance are observed at relatively high temperature T = 160 K. In Fig. 2 of the resistance versus field plots of CC1 and CC2, experimental data exhibit peak and valley type features that seem to be quasi-oscillatory in nature and appear along one field scan (initial) direction only, as highlighted by the color scheme. In order to reproduce the result, the sample needed to be warmed up to room temperature in zero field. An experimental procedure was devised which we found of critical importance for the reproducibility of results. The procedure involved warming the sample to 300 K in zero field, staying at 300 K for 30 minutes before cooling to the measurement temperature $T$ (5 K $\leq$ $T$ $\leq$ 300 K) in zero field, followed by the $R$ vs. $H$ measurement.The peak type features in magnetoresistance were much stronger in CC1 compared to CC2, which indicates that the observed phenomena depends on the thickness of the polycrystalline Cu film . Unlike in the nanoengineered samples, a set of symmetric peaks in resistance is observed  in both field scan directions in non-periodic sample CCR.\cite{Berkowitz}The symmetric peaks in magnetoresistance reside on top of a linear slope, which is also evident in CC1 but absent in CC2. We note that both CCR and CC1 have same Cu film thicknesses while the thickness is reduced by half in CC2. Therefore, a linear slope can arise due to the paramagnetic scattering  of conduction electrons from excessive Cu in these samples. The experimental data of CCR (see supplementary materials)  is interesting in its own right and requires further analysis for complete understanding. However, we will focus on the electrical properties of nanoengineered samples in the rest of the letter.

In order to elucidate the anomalous observation in nanoengineered materials, most notably in CC1, we systematically performed a set of electrical measurements in applied magnetic field and at many temperatures in the range of 5 K $\leq$ $T$ $\leq$ 300 K. In Fig. 3a-h, we have plotted the magnetoresistance $MR$ (= [R($H$)-R(0)]/R(0)) versus $H$ data of CC1 at few characteristic temperatures. The $MR$ data are result of first magnetic field scans (as highlighted by black markers in the color scheme of Fig. 2). These experimental observations were reproduced on three identical but separately fabricated samples of CC1. In Figure 3, we immediately notice a trend in the development of peaks and valleys in the MR data, as the temperature is reduced below 200 K. At $T$ = 200 K, the peak-like structures develop around 0.2 and 0.8 T while the curve sweeps from the positive $MR$ at low field to the negative value at high field. As the temperature is reduced, peaks and valleys become more pronounced and create an oscillatory pattern, which is very pronounced at $T$= 140 K, Fig. 3e. Measurements were also performed in the field-cooled condition for the various selection of cooling field in the range of 0 to 7 T. But no such behavior in MR was observed (see supplementary material).

At first instance, this behavior seems to be reminiscent of Shubnikov de-Hass (SdH) type oscillation.\cite{Xiu} The SdH oscillation represents the quantization of Landau energy levels due to the orbital motion of free electrons in magnetic field, applied perpendicular to the plane of a semiconductor heterostructure. The quantization of Landau levels becomes stronger at higher magnetic field. Therefore, the SdH oscillation is more prominent at higher field, which is opposite to the observation in Fig. 3e. For further decrease in temperature, $T$ $\leq$ 100 K, the slope of $MR$ curve reverses its sweeping course and tend to saturate at higher field values. At $T$ = 100 K, the magnetoresistance is best described by a quadratic function of $H$ at low field values, Fig. 3i. The quadratic increase of resistivity under applied magnetic field is related to the Lorentz force $\textbf{F}$ = q$\textbf{v}$$\times$$\textbf{B}$ on the charge carriers, also termed as $^{'}$ordinary$^{'}$ MR. In addition to this macroscopic effect, a microscopic effect due to the spin-orbit interaction between the electron orbit and the magnetization also plays a significant role in magnetic systems.\cite{Porter} The latter effect, which is purely magnetic in origin, does not explain the quasi-periodic behavior in $MR$ as a function of $H$ in CC1. As such effect would lead to similar experimental observation in CC2 as well, where magnetic dot configuration is identical to CC1. The MR data of CC2 are plotted in Fig. 4a-d. The peak like structures are also visible in the MR data of CC2, perhaps much weaker. Unlike CC1, the MR is negative at $T$$\leq$100 K and does not exhibit the quadratic field dependence at low field. The stark discrepancies in the MR data of CC1 and CC2 clearly suggests that the anomalous observation depends on the number of available charge carriers in the composite system, which can be tuned by varying the thickness of the Cu film.

\begin{figure}
\centering
\includegraphics[width=16cm]{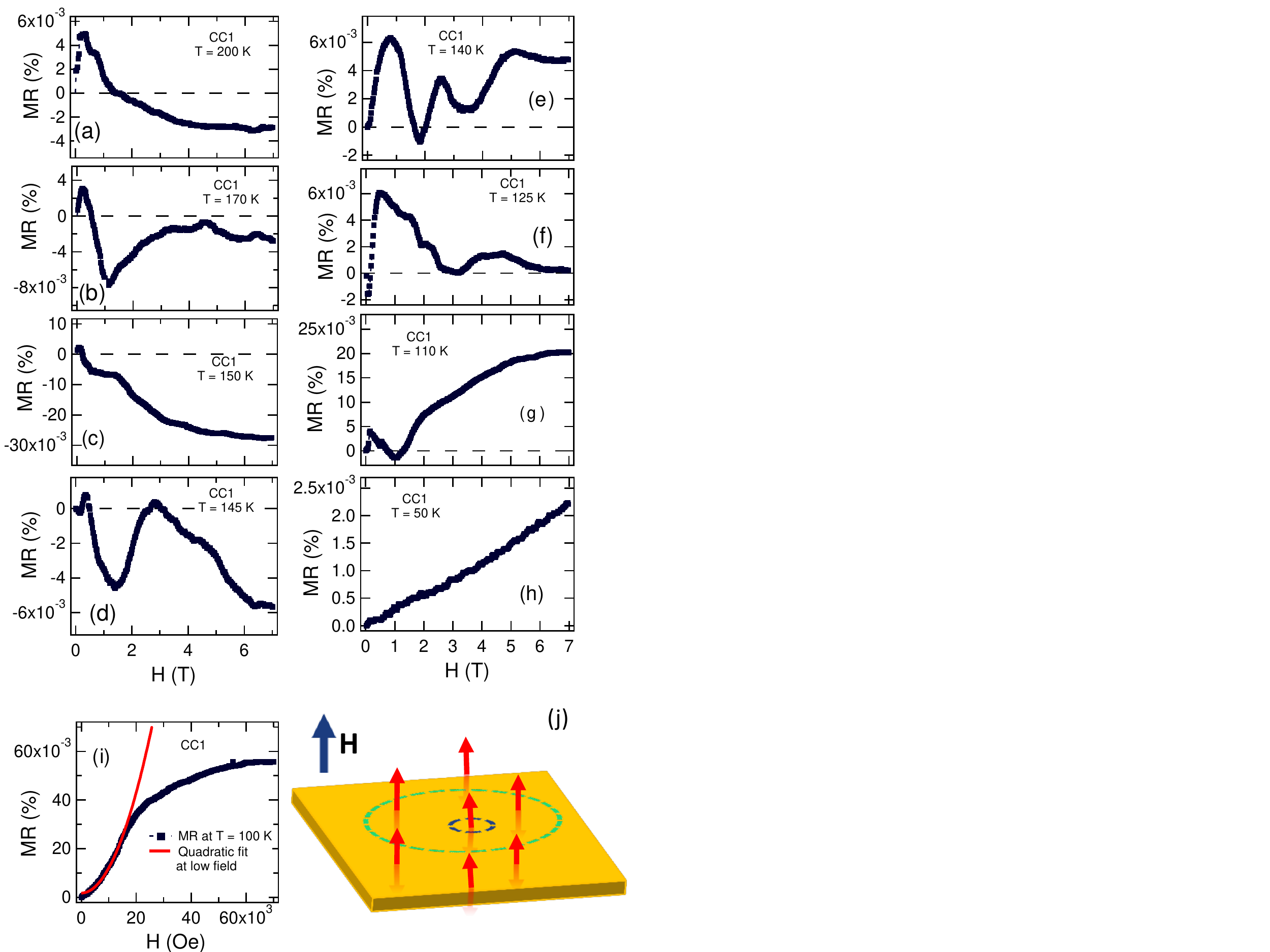} \vspace{-2mm}
\caption{\textbf{Electrical measurements of CC1 at various temperatures}. \textbf{a-h,} $MR$ versus $H$ measurements of CC1 at various temperatures. As we see in these plots, the peak-like structures in $MR$ evolve into anomalous oscillation as the temperature is reduced. \textbf{i,} At $T$ = 100 K, the oscillatory behavior is replaced by a quadratic field dependence at low fields, indicating the Lorentz force effect on charge carriers. \textbf{j,} Schematic description of conduction electrons orbiting the giant moments (red arrows) in different orbits at different fields (see text).
} \vspace{-4mm}
\end{figure}

\begin{figure}
\centering
\includegraphics[width=10.5cm]{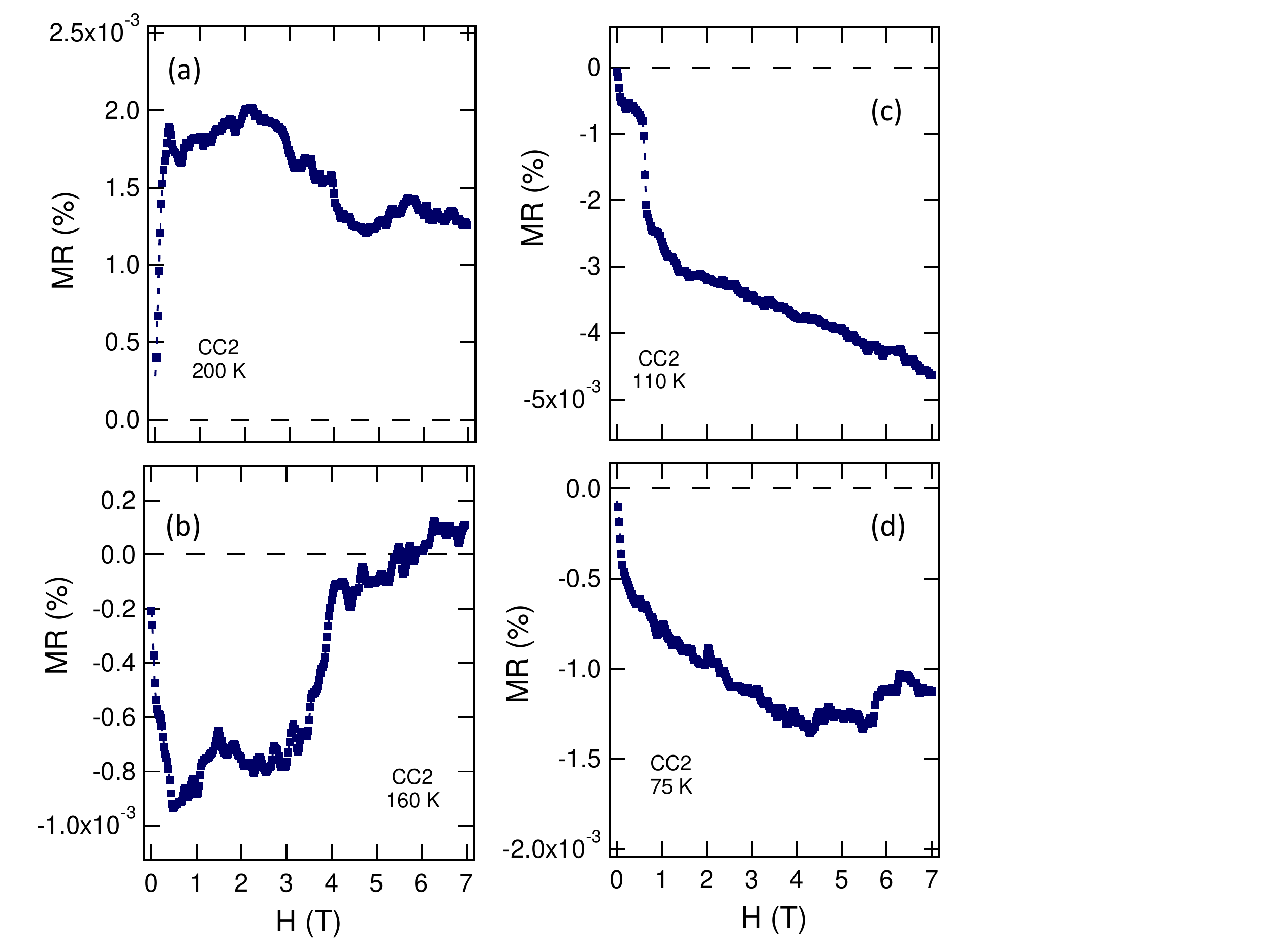} \vspace{-2mm}
\caption{\textbf{Electrical measurements of CC2 at various temperatures and fields}. \textbf{a-d,} $MR$ versus $H$ measurements of CC2 at various temperatures. Very weak peak-like structures in $MR$ are visible in CC2 also.
} \vspace{-4mm}
\end{figure}

In many composite systems of Cu and Co elements in various artificial configurations, such as layered material, the underlying physics is often dictated by the Ruderman-Kittel-Kasuya-Yoshida (RKKY) type oscillatory indirect exchange interaction between Co sites via a thin layer of Cu film, typically ranging from 0.1 - 2 nm.\cite{Parkin,Berkowitz} In the nanoengineered system CC1, the closest separation distance between Co dots is $\simeq$15 nm. A theoretical calculation of effective exchange energy between two Co clusters (of approximately 140 atoms each) immersed in Cu environment suggests that such a mechanism would give rise to a very weak exchange constant $\simeq$ 0.03 meV (equivalent to 0.3 K), at half of the closest separation distance between two magnetic sites, d $\simeq$ 7 nm.\cite{Altbir} Therefore, the oscillatory behavior of magnetoresistance, which occurs at relatively high temperature, cannot be explained by such a weak exchange coupling. The RKKY interaction, however, is strong enough at intra-spin separation inside a Co dot site to cause strong correlation between Co spins and form a giant magnetic moment.\cite{Altbir}

A qualitative explanation of this anomalous MR behavior possibly involves the coupling between the conduction electrons orbital motion and the localized moments in applied magnetic field, akin to the spin-orbit coupling in a ferromagnetic system. At different field values, the conduction electrons orbit the giant localized moments in different radii, encompassing different number of dot sites, to minimize the energy of the system. Since the number of conduction electrons in CC1 is double of that in CC2, the weaker peak like structures in CC2 can be qualitatively understood on this basis. A schematic description of this effect is shown in Fig. 3j. A detail quantitative calculation will require many body approach. The polycrystalline nature of the Cu film will lead to a highly disordered glassy-phase of the system as the mean free path is limited by the grain size of the Cu ($\simeq$ 1 nm). Therefore, reversing the field will require a lot of energy to change the orbital direction of conduction electrons. Such unavailability of energy to the system can limit the quasi-periodic behavior of MR along one field scan direction only. However, the unusual temperature dependence of the MR oscillation is still a mystery and will require further theoretical and experimental research to understand that. It is reasonable to assume that the topography of the system plays an important role in this anomalous observation. These observations suggest a new phenomenon, which can be exploited for the future designs of spin valve and spin sensors for technological applications.

\section{Methods}
The sample fabrication process of nanoengineered materials involve the development of hexagonal nanoporous copolymer templates from a self-assembled diblock copolymer film of thickness 30 nm, average pore diameter $\simeq$ 12 nm and lattice constant of 28 nm on top of a silicon nitride (SiN$_{x}$) substrate.\cite{Russell} For this purpose, a 0.5 $\%$ PS-b-P4VP copolymer solution in toluene/THF (80:20 v/v) was prepared at 65$^{o}$ centigrade and cooled to the room temperature. PS-b-P4VP thin films were spin coated at 2000 rpm for 60 second onto cleaned silicon nitride wafers and dried for 12 hours in vacuum. The samples were solvent annealed in THF/toluene (80:20 v/v) mixture for 12 hours at 23$^{o}$ centigrade and developed in methanol for 20 mins to yield porous templates. The nanoporous copolymer template was used as a mask to deposit cobalt on silicon nitride substrate in locally hexagonal order, forming a periodic array of magnetic dots. The thickness of individual cobalt dot was estimated using atomic microscope micrograph, $\simeq$ 3 nm. After material (cobalt) deposition, the sample was rinsed with toluene to remove the remaining polymer template encapsulated with thin cobalt film, leaving only the magnetic dot array on SiN$_{x}$ substrate. A thin layer of copper film (30 nm) was sputtered on top of magnetic dot array after ionized argon plasma treatment at 50 Watt for 15 minutes. Argon plasma cleaning of the sample helps remove or reduce the native oxide layer of cobalt, thus provides a cleaner, multi-directional contact between the magnetic dots and the polycrystalline Cu film. Electrical measurements were performed in a Quantum Design PPMS, using the built in resistance bridge. The samples gradually became unstable after roughly twenty thermal cycles.

\section{Acknowledgements}  This work was supported in part by the NSF under Agreement No. DMR-0944772 and CHM grant CMMI-1025020. We thank Julie Borchers and Dan Neumann for helpful discussions.

\section{Authors contribution} DKS and MTT envisaged the research idea. SD, DKS, RT carried out the experiments. DKS and MTT prepared the manuscript.

\section{Correspondence} Correspondence and request for materials should be sent to DKS (email: dsingh@nist.gov) and MTT (email: tuominen@physics.umass.edu).

\clearpage

\end{document}